\documentclass[a4paper,11pt]{article}
\usepackage{jinstpub} 
\usepackage{lineno}
\usepackage{tabularx}

\title{\boldmath Rapid cargo verification with cosmic ray muon scattering and absorption tomography}

\author[a,b]{Anzori Sh. Georgadze} 
\affiliation[a]{ Institute for Nuclear Research of the National Academy of Sciences of Ukraine,\\Prospekt Nauky 47, 03680, Kyiv, Ukraine}
\affiliation[b]{Institute of Physics, University of Tartu,\\ W. Ostwaldi 1, 50411, Tartu, Estonia}
\emailAdd{anzori.heorhadze@ut.ee; georgadze@kinr.kiev.ua}

\abstract{Cosmic ray muon tomography is considered a promising method for the non-invasive inspection of shipping containers and trucks. It utilizes highly penetrating cosmic-ray muons and their interactions with various materials to generate three-dimensional  images of large and dense materials, like inter-modal shipping containers, typically not transparent with conventional X-ray radiography techniques. The commonly used methods for imaging with muons are based on muon scattering or absorption-transmission data analysis. Due to large thickness of cargo material in shipping container substantial scattering and absorption occur when muons are passing through cargo. 

One of the key tasks of customs and border security is to verify shipping container declarations to prevent illegal trafficking, and muon tomography could be a viable choice for this task. In this paper, we demonstrate through Monte Carlo simulations using the GEANT4 toolkit that a combined analysis of muon scattering and absorption data can improve the identification of cargo materials compared to using scattering or absorption data alone.
The statistical differences in scattering and absorption data for several cargo materials are quantified. For a particular smuggling scenario where tobacco declared as paper towel rolls, it is demonstrated that the combined analysis can accurately distinguish between tobacco and paper towel rolls with 5.5$\sigma$ accuracy for detector spatial resolution (FWHM) of 0.235 mm,  4.5$\sigma$ for 1.175 mm resolution (FWHM), and 3.9$\sigma$ accuracy for 2.35 mm spatial resolution (FWHM), in a short scanning time of 10 seconds. This rapid detection capability has significant implications for anti-smuggling efforts and cargo inspection.}

\arxivnumber{2407.01020} 
\begin{document}
\maketitle
\flushbottom

\section{Introduction}
Muon tomography is an emerging technique with promising applications in various fields such as non-destructive testing, border security, and archaeology~\cite{bonechi,barnes2023cosmic, borozdin2003, explosives, yifan2018discrimination, lowZ}. 
The principle behind muon scattering tomography is similar to medical X-ray imaging, but instead of using X-rays, it utilizes muons. 
This technology has several advantages, including its ability to penetrate dense materials without causing harm and its capability to provide detailed images of large and massive objects. 
Traditional methods of cargo inspection, such as X-ray scanning, require high-intensity sources of hazardous radiation, which have limited penetration power in dense materials and may not provide a comprehensive view of container or cargo contents.

Initially, the goal of muon tomography was to detect nuclear material smuggling to prevent the threat of nuclear terrorism. Later, muon tomography was proposed for application in border security to scan cargo containers and vehicles for hidden contraband or illicit materials without the need for physical inspection, thereby enhancing security while minimizing disruptions to trade and travel~\cite{schultz2003cosmic, aastrom2016precision, checchia2016review, antonuccio2017muon, pugliatti2014design, morris2013new, preziosi2020tecnomuse, georgadze2023geant4, georgadze2024simulation, georgadze2024,Anzori_Auto}. 
The European project, "Cosmic Ray Tomograph for Identification of Hazardous and Illegal Goods Hidden in Trucks and Sea Containers" (SilentBorder)~\cite{sbwebsite}, focuses on the development and in-situ testing of a high-technology scanner designed for border guards, customs, and law enforcement authorities to inspect shipping containers at border control points.

Muon tomography utilizes cosmic ray muons, which are the secondary particles that come from the extensive atmosphere shower of high-energy cosmic rays from space, mainly protons with a flux about $\approx$ 10000/m\textsuperscript{2}/minute. The distribution of zenith angle $\theta$ is known to follow a cosine-squared law, such that $\textit{I}(\theta)$ = $\textit{I}_0 \cos^2\theta$. Muons are a natural source of radiation that rain down upon the Earth with an average energy of 3–4 GeV. Their mass is approximately 207 times heavier than the mass of an electron ~\cite{tanabashi2018review}.
As these muons pass through matter they undergo multiple Coulomb scattering. The degree of scatter observed is dependent on the Z of material. 

When muons interact with matter, they scatter, and by measuring the angles and intensities of these scattered muons it is possible to create 3D images of the interior of objects or structures. 
In addition, a fraction of cosmic muons can be absorbed by the crossed materials. 
Muon absorption depends on the density and composition of the material. When a muon interacts with matter, it can lose energy through various processes such as ionization, bremsstrahlung, and nuclear interactions. There are two main types of cosmic ray muon imaging techniques: absorption-based and scattering-based. Absorption-based imaging exploits the principle that the flux of muons reaching a particle detector decreases with the density of the material they traverse. Dense materials, such as metals or high-density objects, attenuate or absorb more muons than less dense materials. On the other hand, scattering-based tomography relies on the scattering of muons by the nuclei of the materials they pass through. By analyzing the intensity and angular distribution of muons before and after passing through the sea container or truck, the internal structure and density of cargo can be studied. Some previous studies on the application of muon scattering and absorption methods can be found in the following publications 
~\cite{vanini2019muography, Blanpied, rengifo2024design}. 

In our study we have performed detailed GEANT4 simulations of a muon tomography system that employs  detectors with high spatial resolution based on scintillator technology. To improve the accuracy of verifying the contents of shipping containers, we developed a novel approach that combines muon scattering and absorption data. This analysis enabled significantly better material discrimination compared to analyzing only muon scattering data, thereby decreasing measurement time, which is critical for maintaining high throughput in container verification. 
\section{Muon tomography}
The scattering muon tomography technique calculates the deflection of muons from their straight trajectory due to multiple coulomb interactions, which in turn depend on cargo density and chemical composition. 
The absorption muon tomography technique calculates the fraction of muons that were stopped in cargo. The absorption depends on cargo density and chemical composition. 
Due to the different sizes and loading configurations of the cargo, object detection techniques can be applied to localize and identify the dimensions of the cargo in the reconstructed image of the container.
\begin{figure*}[t]
\begin{minipage}{1.\linewidth}
\centering
\includegraphics[width=0.6\linewidth]{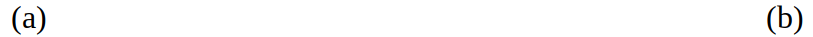}
\vspace{-1.mm}  
\end{minipage}		
\begin{minipage}[t]{0.5\textwidth}
\centering
\includegraphics[width=0.8\textwidth]{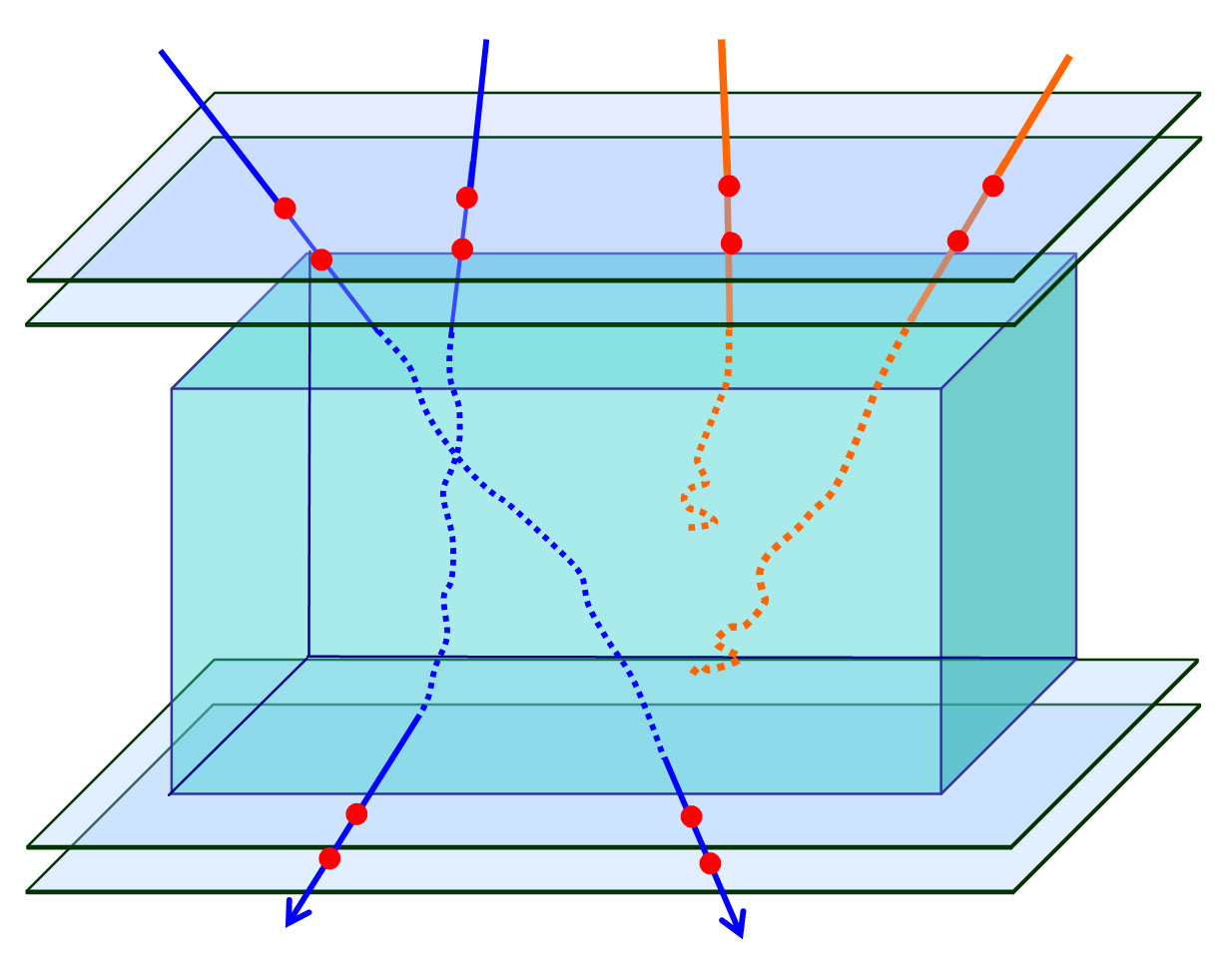}
\end{minipage}
\begin{minipage}[t]{0.5\textwidth}
\centering
\includegraphics[width=0.9\textwidth]{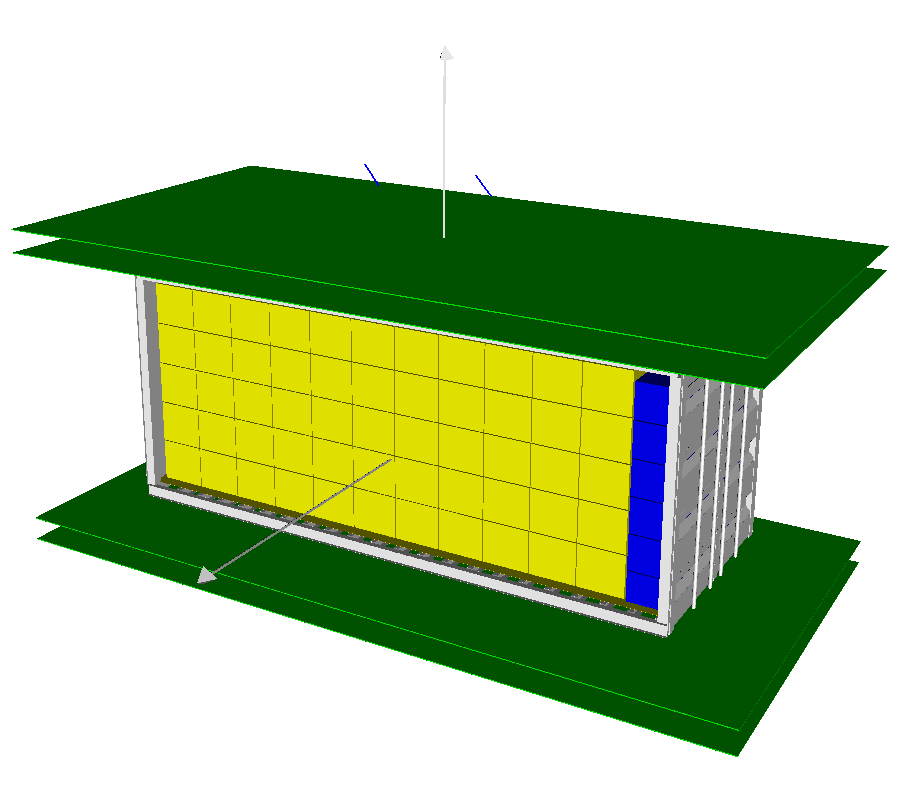}
\end{minipage}
\caption{(a) Visualization of muon scattering tracks (blue) and muon stopped (red) tracks in cargo; (b) muon tomography station (MTS) layout built in GEANT4 simulates a tobacco smuggling scenario. Tobacco fills the entire volume of the container (yellow boxes), and a row of paper towel rolls (blue boxes) is placed between the tobacco and the doors to veil the tobacco. } 
\label{fig:figure1}    
\end{figure*}
Figure~\ref{fig:figure1}a shows examples of scattering muon tracks and stopped muon tracks. On the left two scattered muons that went through the cargo are shown in blue color. The right particles (in red color) are stopped in cargo before reaching the lower tracker detectors. 

\subsection{Muon scattering method}
Cosmic ray muon undergoes multiple coulomb scattering while passing through material. The angular distribution of scattered muon of momentum p is approximately Gaussian, with zero mean and standard deviation given by:
\begin{equation}\label{eq:1}
	\sigma_\theta = \frac{13.6 MeV }{\beta cp}\sqrt{\frac{L}{_{X_{0}}}}(1+0.038)ln\frac{L}{_{X_{0}}})
\end{equation}
where $\beta$ is the ratio between velocity of muon \emph{V} to velocity of light \emph{c}, \emph{X\textsubscript{0}} is the radiation length of the material, \emph{L} is the length of the material traversed. \emph{X\textsubscript{0}} is a material property and depends on the density of the material $\rho$, the atomic mass \emph{A} and the atomic number \emph{Z} and can be expressed as\cite{lynch1991approximations}:
\begin{equation}\label{eq:2}
X_{0} = \frac{716.4g/cm^2}{\rho }\frac{A}{Z(Z+1)ln(287/\sqrt{Z})}
\end{equation}
For material discrimination purposes in muon tomography we define the scattering density ($\lambda$), derived: 
\begin{equation}\label{eq:3}
\lambda(X_{0})= \left(\frac{13.6\;\mbox{MeV}}{p_{0}}\right)^{2}\frac{1}{X_{0}}
\end{equation}
where $p_{0}$ represents the nominal muon momentum, chosen to be 3 GeV/c, and 
$\lambda$ is with units in milliradians$^2$/cm. 
The approximation $\beta$c $\cong$ 1 for muons is used in this equation. The scattering density of a material represents the mean square scattering angle of muons passing through a unit depth of that material. For generating tomographic image scattering angles are accumulated at different locations generating a 3D map of scattering densities.

From equations \ref{eq:2} and \ref{eq:3} one can see that the width of scattering angle distribution inversely depends on the material radiation length which is dependent on the atomic number of the material and material density. For materials with higher atomic numbers Z (high-\textit{Z}) muons will scatter with larger scattering angles. It is therefore possible to characterize the density profile of a large object by measuring muon directions before and after passing through the target object. 
Cargo is often composed of composite materials, and muon scattering is highly dependent on its chemical composition. The radiation length in a mixture or compound may be approximated by~\cite{beringer2012review}:
\begin{equation}\label{ratio}
	\frac{1}{X_{0}}=\sum_{}^{}\frac{\omega_{i}}{X_{i}}
\end{equation}
where $\omega_{i}$ and $X_i$ are the fraction by weight and the radiation length for the \textit{i}-th element.

\subsection{Muon absorption method} 
While passing cargo muons loose their energy due to ionization and have a probability to  be stopped and decay. The attenuation length depends on the density of the matter traversed. 
In the absorption tomography approach, the tracks of muons stopped in the cargo can be reconstructed using an algorithm similar to that described in the publication~\cite{vanini2019muography}.
The muon absorption algorithm analyzes the information regarding the fraction of absorbed muons along the so-called Line-of-Response (LoR), connecting upper and lower muon tracking detectors.
More specifically, the volume containing the cargo material is divided into a collection of voxels (three-dimensional pixels). This algorithm focuses on muons that are detected by the upper detectors but not by the lower detectors due to absorption within the imaging volume.
As defined by the detector geometry, four detectors are used: two upper muon tracking detectors and two lower muon tracking detectors. Scattered muons are identified using hits in both the upper and lower muon tracking detectors. Some muons do not produce hits on the lower detectors because they are either absorbed within the imaging volume or scattered out of the imaging volume. 
For the muons in each LoR, the algorithm calculates the path of the muon through the voxels of the imaging volume. 
This involves determining which voxels the muon ray intersects. The algorithm counts the number of times each voxel is traversed by these muon rays. This count indicates how many muons pass through each voxel but do not exit the imaging volume, suggesting absorption within those voxels.

The reconstructed muon tracks allow for the calculation of the path length $d_{ij}$ of each muon through each voxel \textit{j} within the container. 
The stopping power $S_{j}$ of each voxel represents the energy loss of muons per unit distance as they travel through that voxel.
The total absorption along each muon track is the sum of the stopping powers of all the voxels the muon traverses.
For each LOR, the predicted number of absorbed muons 
$N_{abs,i}$ is calculated based on the path lengths and the stopping powers of the intersected voxels:
\begin{equation}\label{3}
N_{abs,i} =\sum_{j}d_{ij}S_{j}
\end{equation}
The distribution of absorbed muon rays across the voxels allows for the reconstruction of a 3D map of the material’s density and composition within the container. Voxels with higher absorption rates correspond to areas with denser materials. 

Higher-density materials tend to stop more muons, resulting in a higher count of absorbed muon rays in those voxels. Thus, absorbed (stopped) muons provide information about the material density within the voxels, complementing the scattered muon data.

\section{Monte Carlo simulations }
The geometry of a muon tomography station (MST) is shown in figure~\ref{fig:figure1}(b). MST is consisting of two muon tracking modules above and below of the shipping container. Each tracking module includes two position-sensitive detectors, modeled as a plane made of plastic scintillator with a detection efficiency of 100\% and dimension of 8 m $\times $ 4 m $\times $ 1 mm, covering a shipping container. The distance between two position-sensitive detector planes is 10 cm and distance between upper and lower modules is 3 m. 
The simulation of MTS and cargo in sea container is based on GEANT4 toolkit~\cite{AGOSTINELLI2003250}. For generation of cosmic ray muons at sea level we use the Cosmic-Ray Shower Library (CRY)~\cite{hagmann2007cosmic}. 
The origin points of generated muons were sampled from a horizontal plane surface of 10 m $\times$ 10 m. Generated muons are interfaced with GEANT4 to simulate the interaction of muons with the detector, shipping container and cargo. The simulated data samples were produced by generating 5000 datasets for each material. Datasets were simulated by tracking 100000 muons, which corresponds to scanning time of $\approx$10 seconds. Typically, about 10000 muons are detected in Cosmic ray tomography out of 100000 generated. We use the standard physics list for high-energy particle transport named "FTFP\_BERT". Synthetic datasets were analyzed using ROOT data analysis package ~\cite{ROOT}. 
The prototype of tracking detector based on multilayer array of plastic scintillating fibers readout with Silicon Photomultipliers (SiPMs) was constructed and tested~\cite{Anbarjafari}. The obtained spatial resolution in experimental runs was 120 microns (FWHM). To mimic the material discrimination accuracy of detectors with different spatial resolutions, the hit positions of muons produced in the detector plates were smeared using a Gaussian function with FWHM values of 0.235 mm, 1.17 mm, and 2.35 mm.

Contraband materials can be disguised as legal materials in customs declarations to evade detection. Muon Scattering Tomography (MST) can verify the real density and atomic number of the material in the container, discriminate between different products, and allow customs officials to compare the actual contents of a shipping container with what is declared in the customs documentation. According to research~\cite{descalle2006analysis}, the mean cargo density is just under 0.2 g/cm$^3$. As a source of data for the selection of simulation scenarios one can use a PIERS (Port Import/Export Reporting Service) United States import data set. The sheer volume and complexity of the flow of commercial goods make it impractical to model every possible smuggling scenario. In this work, we have simulated scenarios of uniform cargo loading, which are most commonly used in international trade traffic.
For the simulations we considered cargo materials in bulk form. The bulk density of a material is a combination of the density of solid particles, which includes occluded air, the density of the solids themselves, and the interstitial air. 

Table \ref{table:table1} presents the densities of bulk materials used in the modeling taken from~\cite{webtech,bulk}, as well as their evaluated chemical composition. 
The symbols H, O, C and other chemical symbols represent the elements hydrogen, carbon and corresponding elements of the periodic table. The chemical composition of air was taken from the predefined Geant4 Material Database. 
Due to factors like production technologies, regional differences, and processing techniques, it is challenging to obtain precise chemical compositions for bulk materials used in simulations. This variability impacts the accuracy of material differentiation. In practical scenarios wide database created while scanning of containers at border crossing points will be used for precise material characterization.
\begin{table}[t]
\caption{Properties of bulk material used in the simulations.}
\centering
\begin{tabular}{|p{3.3cm}|p{2.1cm}|p{8.3cm}| }
\hline
\bf Bulk material & \bf g/cm$^3$ & \bf Chemical composition (fraction of mass)\\\hline
Paper (towel rolls)     & 0.120 & C$_6$H$_{10}$O$_5$~\cite{2014Chapter2C}              \\ \hline
Tobacco (cigarette)     & 0.190 &  6\%H + 48\%C + 46\%O~\cite{chumsawat2020utilizing}            \\ \hline
Coffee \hspace{15mm}(Roasted (Beans))& 0.368 & 54.2\%C + 35.9\%O + 6.9\%H + 3.1\%N~\cite{campo2020evaluation} \\ \hline
Tea                     & 0.433 & 31.8\%C + 58.2\%O + 2.6\%Mg + 5.7\%K + 1.3\%Ca~\cite{noori2020development}\\ \hline
Polyethylene (Pellets)  & 0.561 & 60\%C$_2$H$_4$ + 40\%air~\cite{collaboration2019book}           \\ \hline
Rice                    & 0.71  & 85\%C$_6$H$_{10}$O$_5$ + 12\%H$_2$O + 8\%CH(NH$_2$)COOH~\cite{rice}     \\ \hline
Sugar                   & 0.800 & C$_{12}$H$_{22}$O$_{11}$~\cite{genova2007monosaccharides}\\\hline
Dry pasta               & 0.950 & 85\%C$_6$H$_{10}$O$_5$ + 15\%CH(NH$_2$)COOH~\cite{pasta,Starch}\\ \hline
Graphite (granules)     & 1.089 &  54.45\%C + 45.55\%air~\cite{collaboration2019book}  \\\hline
Marble (granular)       & 1.282 & CaCO$_3$~\cite{marble} \\ \hline
Cement                  & 1.362 & 61.66\%CaO + 19.83\%SiO$_2$ + 2.32\%Fe$_2$O$_3$ + 4.48\%Al$_2$O$_3$ + 3.14\%MgO + 2.57\%SO$_3$~\cite{OMERSAEED}\\ \hline
Copper (Fines)          & 1.618 & 18\%Cu + 82\%air~\cite{collaboration2019book}\\ \hline
\end{tabular}
\label{table:table1}
\end{table}

\section{Image reconstruction}
Scattering muon tomography is based on the measurement of muon deflections when passing through the cargo. We calculate deflection using reconstruction algorithm based on the Point of Closest Approach (PoCA) method \cite{hoch2009muon} which calculates the closest point of approach between two 3D lines. 
The PoCA algorithm makes the simplified assumption that the muon scattering occurs in a single-point and searches for the point of closest approach between the incident $\textit{v}_{in}$ and outgoing $\textit{v}_{out}$ reconstructed muon track directions. The shortest line segment between the incident and outgoing tracks is estimated by finding the pair of points of closest approach between the tracks. Midpoint of this line segment is considered as a scattering center of the muon. The scattering angle between the two tracks is calculated with the formula~\cite{lynch1991approximations}:
\begin{equation}
	\theta_{scatt} = \arccos(\frac{\vec{v}_{in} \times \vec{v}_{out}}{\left | \upsilon_1 \right |\left | {\upsilon_2} \right | })
\end{equation}
Degree of muon deflection while passing the cargo in shipping container depends on material density and its chemical composition, thus an amount of reconstructed scattering centers using the PoCA approximation can be used for the material identified.

Tracks of incoming and outgoing muons provide information on the scattering centers and scattering angles of muons passing through the imaging volume. 
To obtain best material discrimination efficiency, specific noise-suppression filters must be applied. 
The imprecision of the PoCA algorithm results in a blurring of scattering centers in space, which can be mitigated by using spatial cuts. 
We reduced the image volume to $580 \times 210 \times 210$ cm$^3$ to remove scattering centers created by the walls of the shipping container. To eliminate sporadic scattering centers, the scattering angle is set in the range of 1–100 mrad. Image also is ﬁltered with a median ﬁlter ~\cite{Anzori_Auto} to discards the voxels with uppermost and lowest density, and replaces the voxel density value with the median value computed with the surrounding voxels. The threshold value applied during data processing improves the contrast of the image by removing noise.
\begin{figure*}[b]
	\begin{minipage}[t]{0.5\textwidth}	
		\centering
	\end{minipage}	
	\centering
	\includegraphics[width=0.7\linewidth]{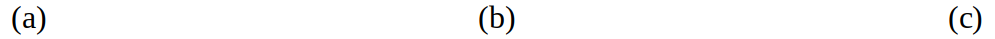}
	\vspace{-1.mm}  
	\begin{minipage}[t]{1\textwidth}
		\includegraphics[width=1.01\textwidth]{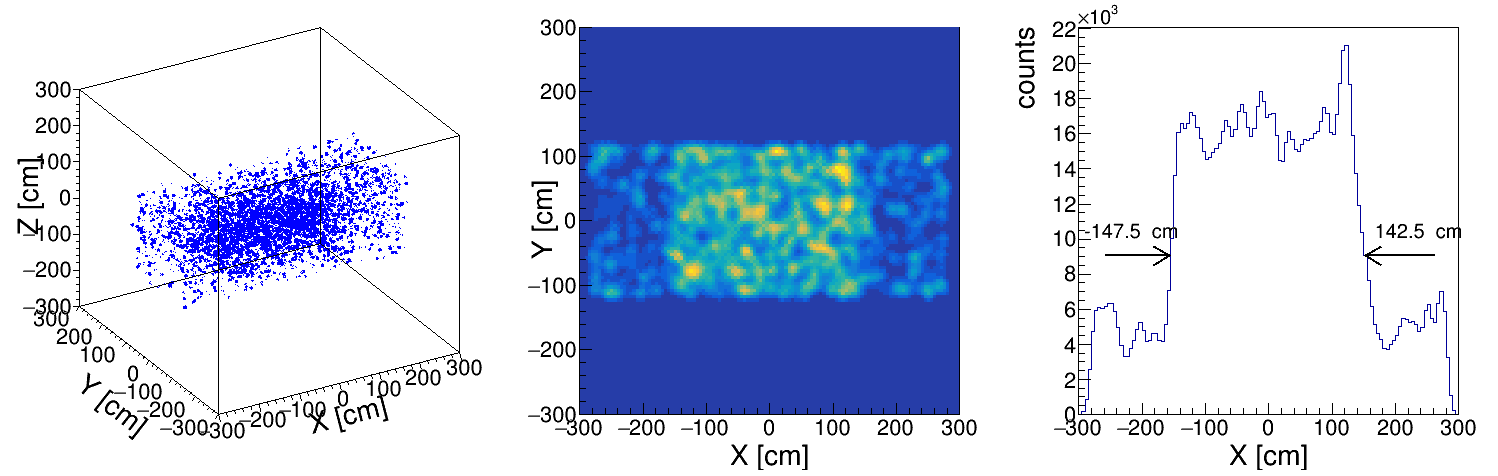}
	\end{minipage}	
\caption{(a) Reconstructed tomographic image of cargo with dimensions $3\times2.3\times2.3$ m$^3$, (b) 
reconstructed X--Y image; (c) projection onto the X-axis of the tomographic image.} 
	\label{fig:figure2}    
\end{figure*}
\section{Results}
The simulation of muon passage through MTS and cargo enables the production of data sets for statistical analysis. Hits on detector planes are recorded for each event, enabling the reconstruction of muon tracks using a ROOT analysis code. 
For the calculation of scattering density and stopping power, cargo dimensions are determined from the reconstructed tomographic image. In the 1D projections of the reconstructed tomographic image obtained from a 10-second scan, the edges of the cargo object can be identified using a ROOT script that calculates the first derivative, as described in publication~\cite{Anzori_Auto}. In figure~\ref{fig:figure2}, tobacco packed into cardboard boxes forms an approximately cubic shape with dimensions $3\times2.3\times2.3$ m$^3$.  The detected cargo edges are found at positions -147.5 cm and 142.5 cm, while the actual positions are -150 cm and 150 cm. Simulated cargo materials were represented by a $5.85\times2.30\times2.30$ m$^{3}$ rectangular volume, almost fully filling the shipping container. In the simulation run, the reconstructed scattering centers and scattering angles together with stopped muon tracks are recorded. For each cargo material, 5,000 datasets were simulated. 

\begin{figure*}[t]
	\begin{minipage}{1.\linewidth}
		\centering
		\includegraphics[width=0.7\linewidth]{figures/abc.png}
		\vspace{-1.mm}
	\end{minipage}
	\begin{minipage}[t]{0.33\textwidth}
		\centering
		\includegraphics[width=1.01\textwidth]{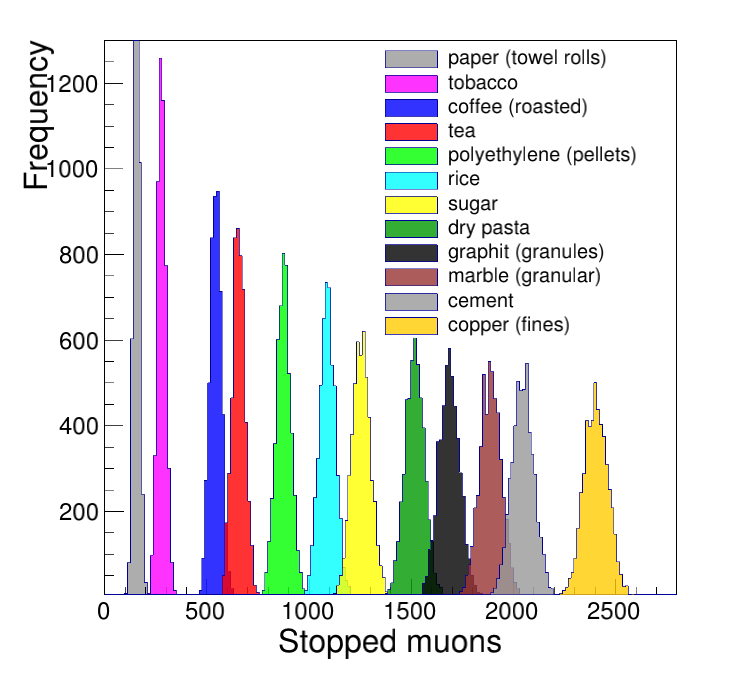}
	\end{minipage}
	\begin{minipage}[t]{0.33\textwidth}
		\centering
		\includegraphics[width=1.01\textwidth]{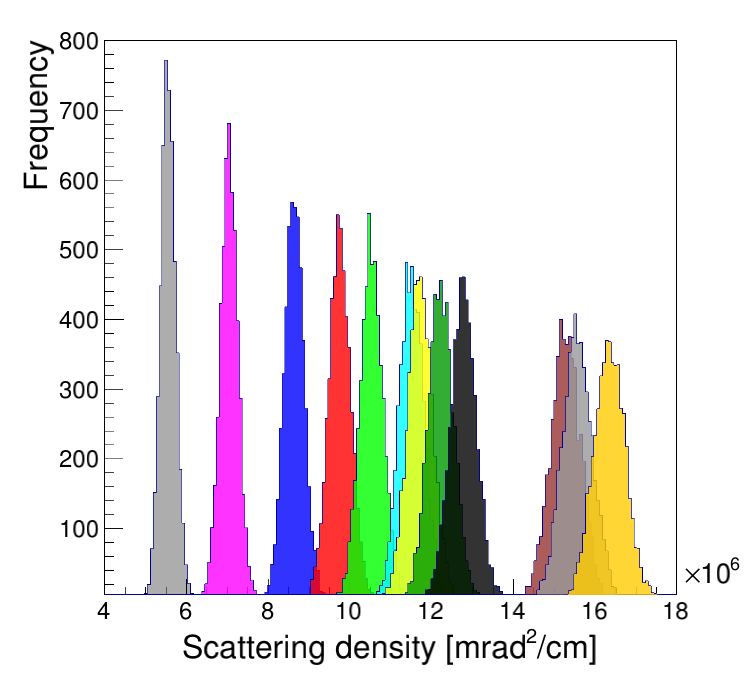}
	\end{minipage}
	\begin{minipage}[t]{0.33\textwidth}
		\centering
		\includegraphics[width=1.01\textwidth]{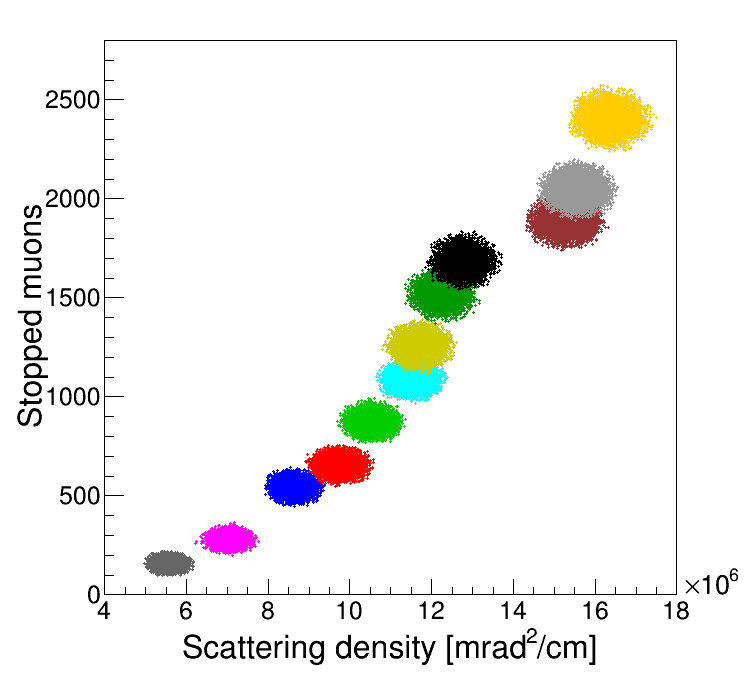} 
	\end{minipage}
\caption{(a) Histogram of the distributions of stopped muons for simulated materials during a 10-second scanning time; (b) histogram of the scattering density distributions for simulated materials; (c) scatter plot of the stopping muon distributions versus scattering density distributions for simulated materials. All simulated distributions are performed for a spatial resolution of 1.17 mm (FWHM). }
\label{fig:scatterplot}
\end{figure*}

Scattering and stopped muon rates distributions for simulated bulk materials are shown of figures~\ref{fig:scatterplot}(a) and (b). In figure~\ref{fig:scatterplot}(c), the scattering density and stopped muons distributions are combined to the two-dimensional histogram for different materials. In 2D space these distributions are well separated and localized in distinct regions, demonstrating better material discrimination ability compared to the discrimination achieved using the 1D presentation of stopped muon and scattering density distributions alone. 

The data show different behaviors in three regions. For the first four types of bulk materials up to tea, the relationship between the scattering and stopping rates is linear, but starting with polyethylene pellets, there is a decrease in the variation of scattering density between materials. This is a result of increased density and total mass of cargo which is resulted in the distortion of muon spectrum (see ~\cite{georgadze2022optimization}). Low-momentum muons (with energies up to around 1 GeV) are more likely to be stopped as they pass through high-density materials like dense organic materials, concrete, or metals. This is because they lose energy more rapidly through ionization and other interactions with the atoms in the material. As the material becomes denser, starting with substances such as polyethylene pellets, the variation in scattering density between different materials decreases. This is due to the distortion of the muon spectrum caused by the absorption of low-momentum muons, resulting in the muon spectrum being increasingly dominated by high-momentum muons. However, high-momentum muons have a lower probability of scattering according to equation \ref{eq:1}, which allows them to penetrate dense cargo with less energy loss, though they provide less potential for differentiation between different materials.  

\begin{figure*}[t]
	\begin{minipage}{1.\linewidth}
		\centering
		\includegraphics[width=0.7\linewidth]{figures/abc.png}
		\vspace{-1.mm}
	\end{minipage}
	\begin{minipage}[t]{0.33\textwidth}
		\centering
		\includegraphics[width=1.0\textwidth]{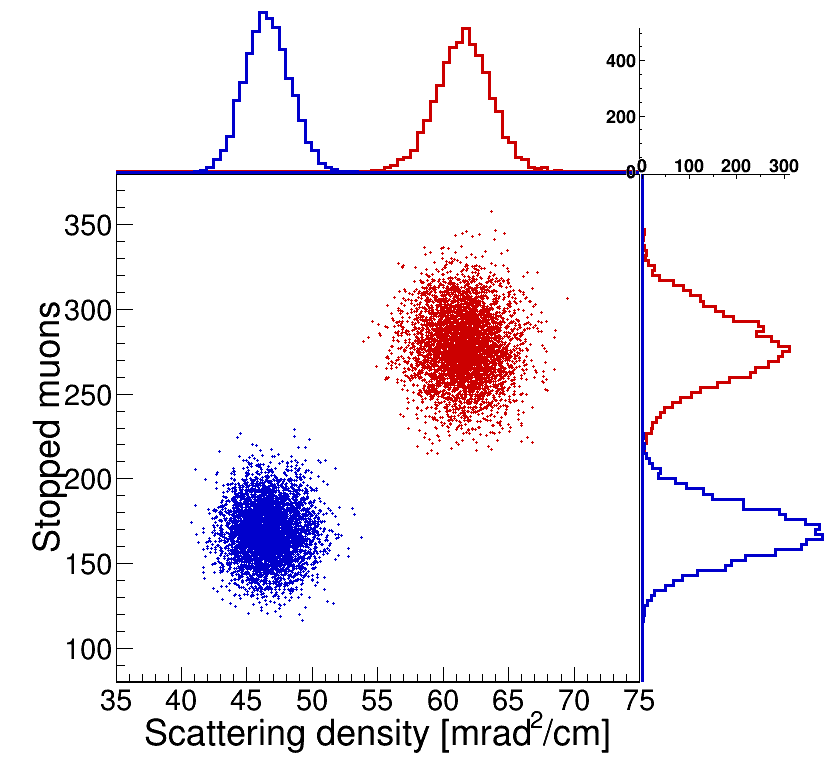}
	\end{minipage}
	\begin{minipage}[t]{0.33\textwidth}
		\centering
		\includegraphics[width=1.0\textwidth]{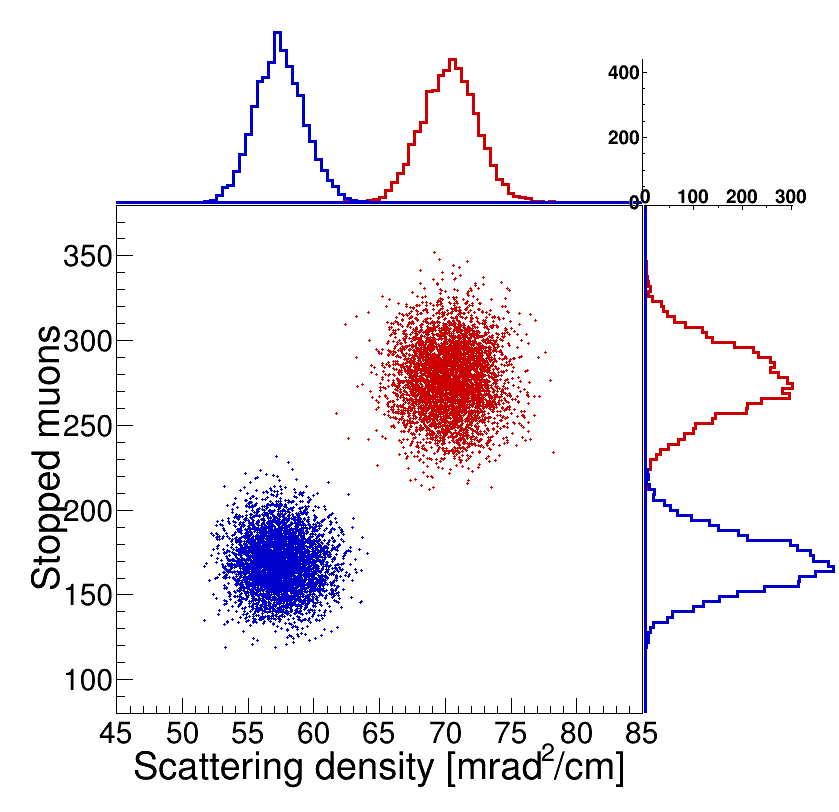}
	\end{minipage}
	\begin{minipage}[t]{0.33\textwidth}
		\centering
		\includegraphics[width=1.0\textwidth]{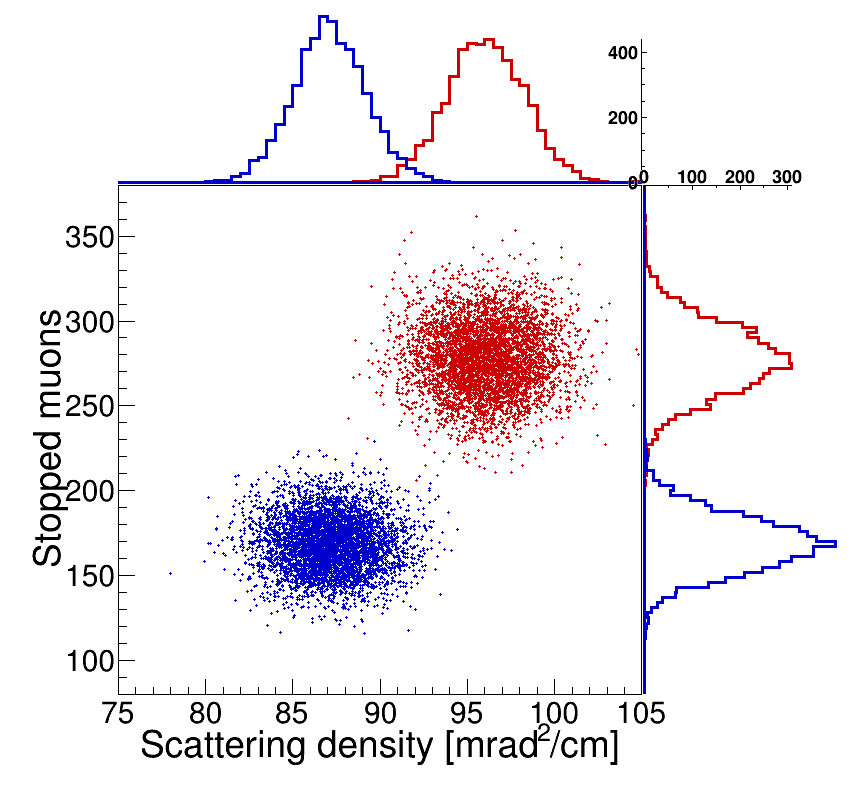} 
	\end{minipage}
\caption{
Scatter plot showing the distributions of scattering density versus stopped muons ratio for paper towel rolls and tobacco. Data points for paper towel rolls are marked in blue, and data points for tobacco are marked in red.
On the top of each 2D histogram we show the 1D histograms of the scattering density distributions for paper towel rolls and tobacco, while on the right side we show the histogram of the corresponding stopped muons distributions.
Panels (a), (b), and (c) of this figure show distributions for detector resolutions of 0.235 mm, 1.17 mm, and 2.35 mm (FWHM), respectively.}
\label{fig:overlapping}
\end{figure*}

To numerically quantify the accuracy of material discrimination, we considered a scenario involving the smuggling of tobacco (cigarettes) declared as paper towel rolls. We simulated datasets for container fully loaded with 5 million cigarettes and container fully loaded with paper towel rolls.  
In figure~\ref{fig:overlapping} the scatter plots of scattering density versus stopped muons distribution are shown for paper towel rolls and tobacco for detector spatial resolutions of 0.235 mm, 1.17 mm, and 2.35 mm (FWHM). Colors indicate group membership of distributions. On the top of each 2D histogram in figure~\ref{fig:overlapping} we show the 1D histograms of scattering data, while on the right side we show the histogram of stopped muons distribution data.
Compared to the scatter versus stopped muons distribution 2D map, the 1D distributions are limited due to the partial overlap of distributions for paper towel rolls and tobacco, making materials discrimination less distinctive. This demonstrates the superiority of 2D data analysis over 1D data analysis.
These plots demonstrate the impact of detector position resolution on material discrimination analysis. 
As can be seen, there is expected degradation of discrimination of accuracy at lower spatial resolution. 

\begin{figure*}
	\begin{minipage}{1.\linewidth}
		\centering
		\includegraphics[width=0.7\linewidth]{figures/abc.png}
		\vspace{-1.mm}  
	\end{minipage}		
	\begin{minipage}[t]{0.33\textwidth}
		\centering
		\includegraphics[width=1.05\textwidth]{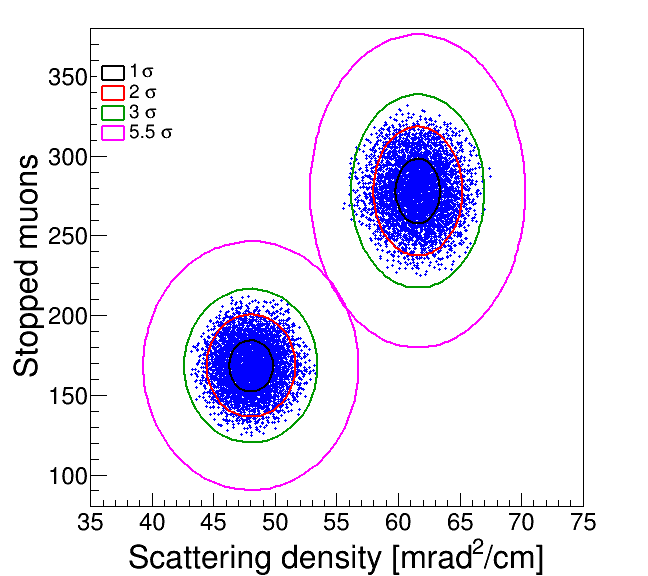}
	\end{minipage}
	\begin{minipage}[t]{0.33\textwidth}
		\centering
		\includegraphics[width=1.05\textwidth]{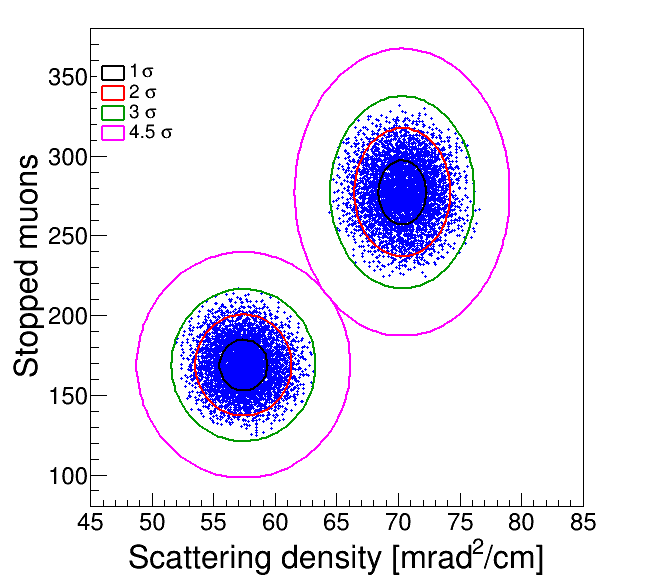} 
	\end{minipage}
	\begin{minipage}[t]{0.33\textwidth}
		\centering
		\includegraphics[width=1.05\textwidth]{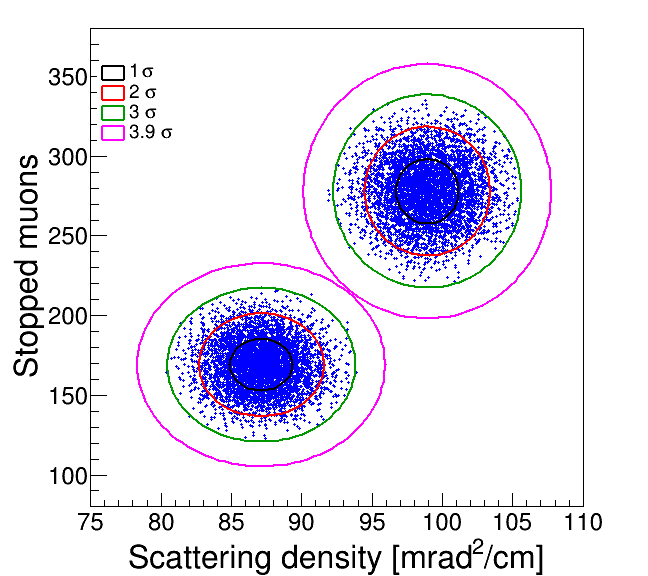}
	\end{minipage}
\caption{(a) 
The scatter distributions of the data samples fitted with two component 2D Gaussian Mixture Model. The black, red and green confidence ellipses for each distribution are set to show 1, 2 and 3  $\sigma$ confidence levels (CL). Panels (a), (b), and (c) of this figure show scatter distributions and confidence ellipses for detector resolutions of 0.235 mm, 1.17 mm, and 2.35 mm (FWHM), respectively. The magenta confidence ellipses show the CLs at which the distributions are discriminated. } \label{fig:ellipse}  
\end{figure*}

For quantitative analysis of discrimination accuracy, we applied the 2D Gaussian Mixture Model (GMM) algorithm, which helps categorize data into groups based on their probability distributions. We initialize a 2D GMM with two components and fit it to the data on the scatter plot (see figure~\ref{fig:ellipse}). The confidence ellipses for each distribution are set to show 1, 2, 3 $\sigma$ CLs and largest confidence ellipse (magenta color) at which distributions are discriminated. As can be seen, the scattering density versus stopping distributions for towel paper and tobacco are accurately discriminated with 5.5 $\sigma$ CL for the detector spatial resolution of 0.235 mm (FWHM), with 4.5 $\sigma$ CL for 1.17 mm (FWHM) and 3.6 $\sigma$ CL for 2.35 mm (FWHM) resolution.
\section{Conclusions}
In this work, we used Monte Carlo simulations based on the GEANT4 package to investigate the ability of cosmic ray muon tomography to perform prompt characterization of shipping container contents and verification of customs declarations. 
We developed a method of combined muon scattering and absorption data analysis to improve the material discrimination ability of cosmic ray muon tomography. This method was applied to analyze simulated datasets for several cargo materials. We demonstrate how integrating scattering and absorption data into a two-dimensional representation enables clearer differentiation between material. The developed approach leverages the combined information from both types of data to enhance the accuracy of material discrimination in muon tomography applications, providing a powerful tool for detecting smuggled goods disguised as other items.
For the specific case of tobacco smuggling, the combined analysis of scatter-absorption data allows for the accurate discrimination between tobacco and paper towel rolls with 5.5 $\sigma$ accuracy for detector spatial resolution (FWHM) of 0.235 mm,  4.5 $\sigma$ for 1.175 mm resolution (FWHM), and 3.9 $\sigma$ which corresponds to 99.95\% discrimination accuracy for 2.35 mm spatial resolution (FWHM) within a 10-second scanning time. 

The ability to rapidly characterize cargo materials comes from the analysis of muon scattering and absorption data for large amounts of cargo materials weighing several tons, resulting in high statistical precision. Thanks to the muon penetration power and the developed method of scatter-absorption combined analysis, cosmic-ray muon tomography can be an effective solution for non-invasive cargo inspections, aiding in the fight against contraband trafficking.

\section{Acknowledgments }
This work was funded by the EU Horizon 2020 Research and Innovation Programme under grant agreement no. 101021812 (“SilentBorder”).

\bibliographystyle{JHEP}
\bibliography{biblio.bib}
\end{document}